\documentclass[aps,prb,twocolumn,twoside]{revtex4}
\usepackage{epsf}
\usepackage[intlimits]{amsmath}
\usepackage{amssymb}
\usepackage{graphicx}
\usepackage{psfrag}
\usepackage{float}

\newcommand{\bscco}{Bi$_2$Sr$_2$CaCu$_2$O$_{8+\delta}$}

\begin{document}

\title{Impurity resonant state in $d$-wave superconductors: 
in favor of a Kondo-like response}

\author{Marijana Kir\'can}
\affiliation{Max-Planck-Institut f\"ur Festk\"orperforschung,
Heisenbergstr.\,1, 70569 Stuttgart, Germany}
\date{\today}

\begin{abstract}
We investigate the impurity resonant state induced
by non-magnetic impurities in $d$-wave cuprate superconductors
in two different impurity models: (i) in a pure potential scattering model within the $T$-matrix approach 
and (ii) in a Bose-Fermi Kondo model within the large-$\mathcal{N}$ formalism.
We modify the superconducting host to resemble 
the nanoscale electronic inhomogeneities seen in the scanning tunneling microscopy (STM) spectra
(i.e., the small- and large-gap domains) and study how this influences the impurity resonance. 
In the pure potential scattering model 
the resonant state is found to be a rather robust feature appearing in all host domains.
On the other hand, in the Kondo-like model 
the impurity resonance can appear in the small-gap regions of the host 
and be completely suppressed in the large-gap regions; 
this is in agreement with the STM experimental
data for Zn-doped \bscco. Our results imply that 
the Kondo spin dynamics of the impurity moment is the origin of the impurity resonant state
in $d$-wave superconductors,
rather than the pure potential scattering.
\end{abstract}


\maketitle


\section{Introduction}

Scanning tunneling microscopy (STM) experiments on 
surfaces of the high temperature cuprate superconductor \bscco~(BSCCO) 
provide a powerful tool for probing strong electronic 
correlations in this material.\cite{hudson,panBSCCO,langBSCCO,uchida}
Since the STM measures a local density of states (LDOS) 
with a high real-space resolution it is widely used 
for studying impurity effects in $d$-wave
superconductors (SC). \cite{balatskyIMP}
Extensive STM studies have been performed on both
magnetic Ni and non-magnetic Zn impurities in BSCCO.
Particularly interesting is the Zn resonant state, 
i.e., a large peak in the differential conductance 
(i.e., in the LDOS) at small but finite energy which is mainly localized
near the Zn site.\cite{yazdani,hudson,panBSCCO}
Despite the fact that there are numerous theoretical
studies dealing with the resonant state induced by
non-magnetic impurities in $d$-wave cuprates, no 
consensus on the theoretical description has been reached
yet.\cite{balatskyIMP}

The Zn$^{2+}$ impurity has spin $S=0$ and it might be natural to describe it
as a delta-function potential scatterer within a potential scattering (PS) 
model.\cite{balatsky,hirschfeld} 
In the PS model the resonance peak can be identified 
with a quasibound state whose energy can be tuned by varying
the strength of the potential. 
However, a low-energy state appears only for large potential values,
i.e., in the unitary limit. 
Moreover, some experimental findings cannot be easily explained within the PS model
like the spatial dependence of the resonance peak which 
shows a global maximum on the Zn site and 
local maxima on the next-nearest neighbor Cu sites.\cite{notefilter}

Nuclear magnetic resonance (NMR) experiments are another
powerful probe of strongly correlated systems like cuprates.
In a series of NMR experiments it has been established that each
non-magnetic impurity (e.g., Zn$^{2+}$ or Li$^{+}$) 
induces a net local $S=1/2$ moment that is distributed 
on the Cu ions in its vicinity.\cite{alloulNMR,mahajanNMR,julienNMR,bobroff}
Furthermore, the effective induced moment interacts with the elementary
excitations of the bulk material giving rise to 
Kondo-like physics which has indeed been 
observed in NMR experiments.\cite{bobroff}
Consequently, one can treat a non-magnetic Zn impurity
in the $d$-wave SC within the Kondo model
and identify the low-energy impurity state in the STM spectra
with the Kondo resonance.\cite{polkovnikovSTM,vojtabullaSTM}
However, it is still not completely settled whether the
potential scattering or the Kondo-like description is more 
appropriate for modeling the non-magnetic impurities in the cuprates
and for explaining all experimental facts.\cite{balatskyIMP}
The main objective of the present study is to
resolve this question.
In principle, Kondo-like behavior 
should be suppressed by an external magnetic field
which would allow to identify the origin of the observed 
impurity STM signal. So far, no such experiments have been
performed and it has been shown that for this purpose
rather high magnetic fields would be needed.\cite{zitzler}

Recently the STM measurements on BSCCO have revealed the presence 
of nanoscale electronic inhomogeneities 
which show up as spatial modulation of both the gap amplitude 
and the LDOS.\cite{howald,panBSCCOa,langBSCCO,mcelroy} 
These electronic inhomogeneities and variations of the LDOS
attracted numerous theoretical and experimental 
studies.\cite{nunner,mcelroySCI,uchida}
Random out-of-plane dopant atoms which change the pair
interaction locally\cite{nunner} and a competing order
like antiferromagnetism\cite{atkinson} have been proposed to be
a possible origin of the LDOS modulations in the cuprates.
Importantly, the STM studies on BSCCO have 
identified two distinct electronic domains: (i)
the domains with a small-gap magnitude and with 
well-developed SC coherence peaks and (ii) the large-gap
domains without SC coherence peaks. 
Remarkably, in the presence of impurities it was
observed that impurity resonances are present only
in the small-gap domains and no impurity resonances can be 
detected in the large-gap domains.\cite{langBSCCO}
Using this experimental fact and motivated by the 
question about the origin of the impurity resonance 
in $d$-wave cuprate SC we investigate how the resonance state is influenced 
by putting the impurity into different domains of the SC host
in both the PS model and in the Kondo model.
To get quantitative results we employ a $T$-matrix 
and a large-$\mathcal{N}$ approach for the PS and the Kondo model,
respectively. We conclude that the Zn impurity resonance in BSCCO 
has predominantly a Kondo-like character.

Very recently a new possibility for the origin of the Zn resonance 
peak in BSCCO has been proposed.\cite{andreev} Namely, in $d$-wave SC
short-ranged off-diagonal ($\tau^x$) impurities generate low-energy 
Andreev states which are then associated with the impurity
resonance seen in the STM spectra. If the usual 
$\tau^z$ impurity scattering is absent, the Andreev resonant 
state exists symmetrically around the Fermi level. 
However, by including a weak $\tau^z$ potential this symmetry is
broken. This gives results which bear closer resemblance 
to the STM data. Below, we will comment on some of these results.

The remainder of the paper is organized as follows.
In Sec.~\ref{sec:model} we introduce the relevant models for
describing a non-magnetic impurity embedded in a $d$-wave superconductor.
In Sec.~\ref{sec:method} the $T$-matrix approach and 
the large-$\mathcal{N}$ formalism are explained. Our
numerical results and a comparison with experimental data are presented
in Sec.~\ref{sec:results}. We conclude the
paper in Sec.~\ref{sec:concl}.


\section{The Model}
\label{sec:model}

To describe a single impurity atom in a superconducting
material we will consider a Hamiltonian consisting of a 
bulk and an impurity part,
$H=H_{\text{bulk}}+H_{\text{imp}}$. 
Both terms are specified below.

\subsection{The Superconducting Host}

As a model for the cuprate superconductor we use the usual BCS Hamiltonian 
\begin{flalign}\label{eq:ham_BCS}
& H_{\text{BCS}}^{\phantom{\dagger}}=
    \sum_{{\bf k}}\Psi_{\bf k}^{\dagger}
    \bigl[(\varepsilon_{\bf k}^{\phantom{\dagger}}-\mu)\tau^z
    +\Delta_{\bf k}^{\phantom{\dagger}}\tau^x\bigr]
    \Psi_{\bf k}^{\phantom{\dagger}},
\end{flalign}
where $\Psi_{\bf k}^{\dagger}=
(c_{{\bf k}\uparrow}^\dagger\,\,\,
c_{{-\bf k}\downarrow}^{\phantom{\dagger}})$
is a Nambu spinor with momentum ${\bf k}=(k_x,k_y)$, 
$c_{{\bf k}\sigma}^\dagger$ ($c_{{\bf k}\sigma}^{\phantom{\dagger}}$)
creates (annihilates) an electron with momentum ${\bf k}$ and spin $\sigma$,
and the three Pauli matrices in particle-hole space are denoted by $\tau^{x,y,z}$. 
We use a tight-binding dispersion of the form  
\begin{flalign}
&  \varepsilon_{\bf k}^{\phantom{\dagger}}=
   \frac{t_1}{2}\bigl(\cos k_x+\cos k_y\bigr)+t_2\cos k_x\cos k_y
   +\frac{t_3}{2}\bigl(\cos 2k_x+\nonumber\\  
&  \quad\quad+\cos 2k_y\bigr)+\frac{t_4}{2}\bigl(\cos 2k_x\cos k_y+
   \cos k_x\cos 2k_y\bigr)+\nonumber\\
&  \quad\quad+t_5\cos 2k_x\cos 2k_y,
\end{flalign}
with values for the hopping elements that fit the ARPES data:\cite{normanBSCCO}
$t_1=-0.5951$\,eV,
$t_2=0.1636$\,eV,
$t_3=-0.0519$\,eV,
$t_4=-0.1117$\,eV,
$t_5=0.0510$\,eV, and
$\mu=-0.1305$\,eV for optimal doping ($p=19.5\,\%$).
The $d$-wave pairing function is given by
$\Delta_{\bf k}=\Delta_0\bigl(\cos k_x-\cos k_y\bigr)/2$
where $\Delta_0=40$\,meV.
In order to model the nanoscale electronic inhomogeneities 
seen by the STM (i.e., the small- and large-gap domains) 
it is necessary to modify the pure BCS state; 
this will be discussed in 
Sec.~\ref{sec:ldos}.

\subsection{The Impurity Hamiltonian}

We consider an effective impurity Hamiltonian which consists of 
a potential scattering and a magnetic term, 
$H_{\text{imp}}=H_{\text{pot}}+H_{\text{K}}$. 
The potential scattering term describes electrostatic interactions 
between the impurity atom and the conduction electrons 
whereas the magnetic term represents an impurity moment embedded in the SC host.

\subsubsection{Potential Scattering Model}

We assume that the potential scattering is completely localized at the impurity site
${\bf r}_0=(0,0)$, i.e.,
$V({\bf r})=V_0\,\delta({\bf r}-{\bf r}_0)$. 
Consequently, the scattering occurs only in the isotropic $s$-wave 
channel and the scattering term is given by
$H_{\text{pot}}^{\phantom{\dagger}}=
    V_0^{\phantom{\dagger}}
    \sum_{\sigma}c_{0\sigma}^{\dagger}
    c_{0\sigma}^{\phantom{\dagger}}$,
where $c_{0\sigma}^{\phantom{\dagger}}=N^{-1/2}\sum_{\bf k}c_{{\bf k}\sigma}$,
and $N$ is the number of sites.
In the PS model the bulk Hamiltonian is defined as 
$H_{\text{bulk}}=H_{\text{BCS}}$.

\subsubsection{Bose-Fermi Kondo Model} 

To specify the magnetic term in the impurity Hamiltonian we use the so-called
Bose-Fermi Kondo model.\cite{bfkSi,BFK1,BFK2,mvIMP}
It describes a spin-$1/2$ impurity, ${\bf S}_{\text{imp}}$, 
coupled both to the electrons of the conduction band and to a bosonic bath
representing bosonic collective excitations in the host material.
(Apart from its application to the
impurity moments in the cuprates,\cite{BFK1} 
the Bose-Fermi Kondo model has attracted considerable 
interest in the context of local quantum criticality within
an extended dynamical mean-field theory,\cite{edmft}
and also in certain mesoscopic systems like 
a quantum dot coupled to ferromagnetic leads.\cite{set})

In the $d$-wave cuprate SC it is reasonable to separate the low-energy 
degrees of freedom into {\it fermionic} Bogoliubov quasiparticles around
the nodal points, 
and into {\it bosonic} antiferromagnetic spin-1 collective fluctuations near the 
ordering wave-vector ${\bf Q}=(\pi,\pi)$.\cite{bkVBS,BFK1,BFK2} 
The effective impurity moment induced by non-magnetic impurity 
interacts with both the fermionic and the bosonic bulk excitations.
Therefore, a non-magnetic impurity
in cuprates is expected to be well described by the Bose-Fermi Kondo 
model\cite{BFK1}
\begin{flalign}\label{eq:kondo_ham}
& H_{\text{K}}^{\phantom{\dagger}}=
    J_{\text{K}}^{\phantom{\dagger}}\,
    {\bf S}_{\text{imp}}^{\phantom{\dagger}}\cdot 
    {\bf s}_0^{\phantom{\dagger}}+
    \gamma_0^{\phantom{\dagger}}\,
    {\bf S}_{\text{imp}}^{\phantom{\dagger}}
    \cdot {\boldsymbol \phi}_0^{\phantom{\dagger}}.
\end{flalign}
Here, ${\bf s}_0^{\phantom{\dagger}}=
1/(2N)\sum_{{\bf kk'}\sigma\sigma'}
c_{{\bf k}\sigma}^{\dagger}
{\boldsymbol \sigma}_{\sigma\sigma'}^{\phantom{\dagger}}
c_{{\bf k'}\sigma'}^{\phantom{\dagger}}$ 
is the conduction electron spin operator at the impurity site,
${\boldsymbol \sigma}$ are the Pauli matrices in spin space,
and $J_{\text{K}}$ is the usual Kondo coupling.
The impurity moment is coupled to the bosonic 
bath via a coupling constant $\gamma_0$. The local
field ${\boldsymbol \phi}_{0}$
represents the local orientation of the antiferromagnetic order parameter
and it is given by
${ \phi}_{0\gamma}^{\phantom{\dagger}}=
\sum_{\bf q}(b^{\phantom{\dagger}}_{{\bf q}\gamma}+
b^{\dagger}_{-{\bf q}\gamma})\sqrt{2JA_{\bf q}/\omega_{\bf q}}$,
where $J$ is the bulk exchange constant,
$A_{\bf q}$ is a dimensionless function containing
the geometry of the system, and $\gamma=x,y,z$.\cite{bkVBS}
The simplest bosonic bath consists of free vector bosons,
$H_{\text{bos}}^{\phantom{\dagger}}=
\sum_{{\bf q}\gamma}\omega_{\bf q}^{\phantom{\dagger}}
b_{{\bf q}\gamma}^{\dagger}b_{{\bf q}\gamma}^{\phantom{\dagger}}$,
with the dispersion $\omega_{\bf q}^2=m^2+c^2{\bf q}^2$,
where the momentum ${\bf q}$ is measured relative to the ordering wave-vector
${\bf Q}$ and $c$ is the velocity of the spin excitations.\cite{bkVBS}
At zero temperature, the effective bosonic mass $m$ is equivalent to
the bulk spin gap $\Delta_{\text{s}}$ 
(as seen in neutron scattering experiments\cite{fong}). 
A spin gap $\Delta_{\text{s}}=0$
corresponds to a bulk quantum-critical point controlling a transition
between a quantum paramagnet ($\Delta_{\text{s}}>0$) 
and the antiferromagnetically ordered phase.\cite{bkVBS}

In the present study the Kondo Hamiltonian \eqref{eq:kondo_ham} 
describes the physics of the so-called pseudogap (i.e., softgap)
Bose-Fermi Kondo model.\cite{BFK1,BFK2}
There, a fermionic LDOS follows a power-law close to the Fermi energy,
$\rho_c^0(\omega)\sim|\omega|^r$, where $\omega$ is measured from the Fermi level.
Clearly, the $d$-wave superconductor has $r=1$.
In addition, at zero temperature and for $\Delta_{\text{s}}=0$,
the bosonic LDOS also follows a power-law, 
$\rho_{\phi}^0\sim\text{sgn}(\omega)|\omega|^{1-\epsilon}$, where
$\epsilon=3-d$, and $d$ is the dimension of the system 
(for cuprates $d=2$).
It is important to note that for $\gamma_0=0$ the Bose-Fermi Kondo model
\eqref{eq:kondo_ham} reduces to the extensively studied
fermionic pseudogap Kondo 
model.\cite{withofffradkin,cassanellofradkin,ingersent,vojtabullaSTM}
Depending on the value of $r$ and 
the presence or absence of particle-hole symmetry,
the fermionic pseudogap Kondo model exhibits a quantum phase 
transition between a phase 
with Kondo screening for $J_{\text{K}}>J_{\text{Kc}}$,
and a phase where the magnetic moment is free (i.e.,
decoupled from the bath) for $J_{\text{K}}<J_{\text{Kc}}$.
It is also known that the presence of an additional
bosonic bath for $\gamma_0\neq0$,
leads to a suppression of  the Kondo screening.\cite{bfkSi,BFK1,BFK2}
Note that in the present analysis we will use the Bose-Fermi Kondo model 
with the bulk Hamiltonian 
$H_{\text{bulk}}=H_{\text{BCS}}+H_{\text{bos}}$.\cite{notecoupling}

To describe non-magnetic impurities like Zn (or Li) in the 
copper oxide planes it is appropriate to use the so-called extended 
magnetic impurity.\cite{polkovnikovSTM, vojtabullaSTM}
NMR experiments have shown that the effective moment induced
by a non-magnetic impurity is spatially distributed around the 
impurity site, ${\bf r}_0$, and mainly resides on the four nearest 
neighbor copper sites.\cite{julienNMR} 
Therefore, the Kondo model with an extended magnetic impurity
is required in order to explain both the STM 
and NMR measurements on cuprates with non-magnetic 
impurities.\cite{polkovnikovSTM, vojtabullaSTM}
In the present calculation we consider a four-site
Kondo model in $d$-wave superconductors where the effective magnetic moment
is coupled to the four Cu sites ${\bf s}=(\pm 1,0),(0,\pm 1)$
adjacent to the impurity. One should bare in mind that
in this four-site model the effective LDOS of the fermionic bath
seen by the impurity changes; this will be shortly discussed
in Sec.~\ref{sec:fermk}.


\section{Method}
\label{sec:method}

We start with the definition of the conduction electron
Green's function in the absence of impurities, which is given by
$\underline{\mathcal{G}}_{\,c}^0(\tau,{\bf k})=
-\langle T_\tau^{\phantom{\dagger}}
\Psi_{\bf k}^{\phantom{\dagger}}(\tau)\Psi_{\bf k}^\dagger(0)\rangle$.
(In our notation all underlined symbols denote matrices.) 
Its Fourier transform is given by
$\underline{\mathcal{G}}_{\,c}^0(i\nu_n,{\bf k})=\int_0^\beta
 \text{d}\tau\,\underline{\mathcal{G}}_{\,c}^0(\tau,{\bf k})\,\text{e}^{i\nu_n\tau}$,
 where $\nu_n=(2n+1)\pi T$ represents the fermionic Matsubara frequencies
 and $\beta$ is the inverse temperature.
 Note that the bare Green's function in real space is
 simply given by $\underline{\mathcal{G}}_{\,c}^0(i\nu_n,{\bf r})=
 N^{-1}\sum_{\bf k}\text{e}^{i{\bf k}\cdot{\bf r}}
[i\nu_n-(\varepsilon_{\bf k}-\mu)\tau^z-\Delta_{\bf k}\tau^x]^{-1}$.

\subsection{$T$-matrix Approach}

The single scatterer problem, $H_{\text{pot}}$, 
can be solved exactly with the usual 
result for the perturbed Green's function\cite{balatskyIMP}
\begin{flalign}\label{eq:tmat_green}
& \underline{\mathcal{G}}_{\,c}(i\nu_n,{\bf r},{\bf r'})=
    \underline{\mathcal{G}}_{\,c}^0(i\nu_n,{\bf r}-{\bf r'})+\nonumber\\
&  \hspace{10mm}
    +\underline{\mathcal{G}}_{\,c}^0(i\nu_n,{\bf r}-{\bf r}_0)\,
    \underline{T}(i\nu_n)\,
    \underline{\mathcal{G}}_{\,c}^0(i\nu_n,{\bf r}_0-{\bf r'}),
\end{flalign}
where the $T$-matrix is given by 
\begin{flalign}
& \underline{T}(i\nu_n)=V_0\,\tau^z
\bigl[\underline{1}-V_0\,\underline{\mathcal{G}}_{\,c}^0
(i\nu_n,{\bf r}_0)\,\tau^z\bigr]^{-1}.
\end{flalign}
The $T$-matrix depends on the strength of the scattering potential and on the 
local conduction electron Green's function,
$\underline{\mathcal{G}}_{\,c}^0(i\nu_n,{\bf r}_0)$.
It is important to note that this local Green's function 
is the only input quantity representing the bulk material;
it differs significantly in the small-gap and the 
large-gap regions of the SC host 
(more details are given in Sec.~\ref{sec:ldos}).

\subsection{Large-$\mathcal{N}$ Formalism}

In order to obtain quantitative results in the Bose-Fermi Kondo model
we use the large-$\mathcal{N}$ formalism.
The large-$\mathcal{N}$ theory can be constructed by generalizing 
the spin symmetry of the impurity moment, the conduction electrons, and
the vector bosons from SU(2) to SU($\mathcal{N}$). 
An antisymmetric representation of 
the impurity spin is given by
${\bf S}_{\text{imp}}^{\mu\mu'}=
f_\mu^\dagger f_{\mu'}^{\phantom{\dagger}}
-q_0^{\phantom{\dagger}}\delta_{\mu\mu'}^{\phantom{\dagger}}$
with the constraint $\sum_{\mu}f_\mu^\dagger f_{\mu}^{\phantom{\dagger}}
=q_0^{\phantom{\dagger}}\mathcal{N}$, 
where $f_\mu^\dagger$ ($f_\mu^{\phantom{\dagger}}$) 
creates (annihilates) auxiliary fermions with $\mu=1,\ldots,\mathcal{N}$ spin flavors.
The constraint is enforced by introducing a Lagrange multiplier $\lambda_0$.
The generalization of the spin symmetry leads to $\mathcal{N}^2-1$ 
triplet bosons denoted by $b_{{\bf q}\mu\mu'}$.\cite{bkVBS} 
In what follows we consider only the particle-hole symmetric case for 
which $q_0=1/2$. Therefore, in the large-$\mathcal{N}$ formalism the 
impurity Hamiltonian \eqref{eq:kondo_ham} is given by
\begin{flalign}\label{eq:kondo_hamN}
& H_{\text{K}}=\frac{J_{\text{K}}}{\mathcal{N}}
\sum_{\mu\mu'}f_\mu^\dagger f_{\mu'}^{\phantom{\dagger}}
c_{0\mu}^\dagger c_{0\mu'}^{\phantom{\dagger}}+
\frac{\gamma_0}{\sqrt{\mathcal{N}}}
\sum_{\mu\mu'}f_\mu^\dagger f_{\mu'}^{\phantom{\dagger}}
\phi_{0\mu\mu'}^{\phantom{\dagger}}+\nonumber\\
& \quad\quad
   +\lambda_0\Bigl(\sum_{\mu}f_\mu^\dagger f_\mu^{\phantom{\dagger}}
     -q_0^{\phantom{\dagger}}\mathcal{N}\Bigr).
\end{flalign}
It is important to emphasize that in the limit $\mathcal{N}\to\infty$
physics of both the fermionic Kondo model
and the Bose Kondo model is controlled by
a corresponding saddle point.\cite{readnewns,bkVBS} 

We proceed with defining the pseudo-fermion Green's function 
 $\mathcal{G}_f^0(\tau)=-\langle T_\tau f(\tau)f^\dagger(0)\rangle$
 and its Fourier transform 
 $\mathcal{G}_f^0(i\nu_n)=(i\nu_n-\lambda_0)^{-1}$.
We drop the spin indices since the SU($\mathcal{N}$) symmetry is preserved.
The full pseudo-fermion Green's function is determined from the 
Dyson equation
\begin{flalign}\label{eq:fullGf}
& \mathcal{G}_f^{-1}(i\nu_n)=
   \bigl[\mathcal{G}_f^0(i\nu_n)\bigr]^{-1}-
   \Sigma_f(i\nu_n),
\end{flalign}
where the self-energy $\Sigma_f(i\nu_n)$ has to be calculated.
In addition, the Lagrange multiplier, $\lambda_0$, can be determined
by employing the occupation constraint for the pseudo-fermions, which yields
\begin{flalign}\label{eq:half_sc}
& \frac{1}{2}\equiv q_0=\langle f^\dagger f\rangle,
    \hspace{5mm}\text{where}\hspace{5mm}
    \langle f^\dagger f\rangle=
    \mathcal{G}_{f}(\tau=0^-).
\end{flalign}
In order to determine $\Sigma_f(i\nu_n)$ in the Bose-Fermi
Kondo model we first briefly present the known
results for two limiting cases, namely for (i) the fermionic 
Kondo, $\gamma_0=0$, and (ii) the Bose Kondo model, $J_{\text{K}}=0$.

\subsubsection{Fermionic Kondo Model: $\gamma_0=0$}
\label{sec:fermk}

For $\gamma_0=0$ by using a simple decoupling approximation one can obtain
the non-interacting form of the Kondo Hamiltonian \eqref{eq:kondo_hamN},
i.e., $H_{\text{K}}^{\phantom{\dagger}}=\sum_{\mu}\bigl[
-(v f_{\mu}^\dagger c_{0\mu}^{\phantom{\dagger}}+\text{h.c.})+
\lambda_0 f_\mu^\dagger f_\mu^{\phantom{\dagger}}\bigr]$
 with the self-consistently determined condensation amplitude 
 $v=J_{\text{K}}^{\phantom{\dagger}}/\mathcal{N}
 \sum_{\mu}\langle c_{0\mu}^\dagger f_{\mu}^{\phantom{\dagger}}\rangle$.
This scheme, the so-called slave-boson mean-field approximation,
becomes exact in the limit $\mathcal{N}\to\infty$ when the physics is dominated 
by a static saddle point. The known result for the pseudo-fermion self-energy 
is given by\cite{mvIMP}
\begin{flalign}\label{eq:sigmaf_fer}
& \Sigma_f^{(\text{f})}(i\nu_n)=v^2\,\mathcal{G}_{c}^0(i\nu_n,{\bf r}_0),
\end{flalign}
where $\mathcal{G}_{c}^0(i\nu_n,{\bf r}_0)=
\text{Tr}\bigl[\underline{\mathcal{G}}_{\,c}^0(i\nu_n,{\bf r}_0)
(\underline{1}+\tau^z)/2\bigr]$ is the local Green's function of the fermionic bath; 
for $V_0\neq0$ one should use \eqref{eq:tmat_green} instead.
The condensation amplitude, $v$, measures the interaction 
between the conduction electrons and the impurity spin. 
The non-zero value of $v$ signals Kondo screening 
whereas for $v=0$ the moment is unscreened (free). 
$v$ can be determined by calculating the corresponding expectation
value of the bosonic operator $c^\dagger_0 f$ which is related 
to the mixed propagator 
$\mathcal{G}_{fc}(\tau)=-\langle T_\tau f(\tau)c_{0}^\dagger(0)\rangle$.
Hence,
\begin{flalign}\label{eq:v_sc}
& v=J_{\text{K}}^{\phantom{\dagger}}
    \langle c^\dagger_0 f\rangle,
    \hspace{5mm}\text{where}\hspace{5mm}
    \langle c^\dagger_0 f\rangle=
    \mathcal{G}_{fc}(\tau=0^-).
\end{flalign}
A simple diagrammatic analysis leads to 
\begin{flalign}\label{eq:fullGfc}
& \mathcal{G}_{fc}(i\nu_n)=-v\,
   \mathcal{G}_f(i\nu_n)\,\mathcal{G}_{c}^0(i\nu_n,{\bf r}_0).
\end{flalign}
The factor $v$ arises due to the fact that no bare mixed propagator  
exists and the minus sign as a result of the anticommutation relations. 
A non-zero condensation amplitude, $v$, at $T=0$, 
decreases with increasing temperature and 
vanishes continuously at the so-called Kondo temperature, 
$T_{\text{K}}$.

The equations \eqref{eq:fullGf}-\eqref{eq:fullGfc} 
are a self-consistent system of equations for 
determining the full $f$-propagator, the condensation 
amplitude, $v$, and the Lagrange multiplier, 
$\lambda_0$. 
These equations can be solved for a given Kondo coupling, $J_{\text{K}}$,  and a local
conduction Green's function at the impurity site.
From the full $f$-propagator we can easily obtain the pseudo-fermion spectral function,
$\rho_f(\omega)=-\text{Im}\,\mathcal{G}_f(\omega+i0^+)/\pi$.
A trivial solution of the self-consistent system of equations is given by 
$v=0$ and $\lambda_0=0$, which corresponds to an impurity spin which is
completely decoupled from the conduction electrons. 
At high enough temperature this is the only solution.
With decreasing temperature another non-trivial solution with $v\neq0$ can appear. 
The non-trivial solution signaling the Kondo screening arises only 
below the Kondo temperature and for $J_{\text{K}}>J_{\text{Kc}}$.
Note that $T_{\text{K}}$ can be determined 
for a given $J_{\text{K}}$ using \eqref{eq:v_sc} in the limit $v\to0$, $\lambda_0\to0$;
at zero temperature we use \eqref{eq:v_sc} in the same
limit to calculate the critical value of the Kondo 
coupling, $J_{\text{Kc}}$.

It is known that the slave-boson mean-field method has numerous 
artifacts at finite temperatures or 
in the vicinity of the quantum critical point for $\mathcal{N}=2$ 
which is the case of physical interest. 
Nevertheless, this approach has been extensively used in the context of 
the pseudogap Kondo model\cite{withofffradkin,cassanellofradkin,polkovnikov}
and is known to capture the qualitative physics of the Kondo screened phase.
                          
We emphasize that in the four-site Kondo model, which we use in the present study, 
the LDOS seen by the impurity is given by the effective propagator
$\sum_{{\bf s},{\bf s'}}\varphi_{\bf s}\varphi_{\bf s'}
\text{Tr}\bigl[\underline{\mathcal{G}}_{\,c}(i\nu_n,{\bf s},{\bf s'})
(\underline{1}+\tau^z)/2\bigr]$. 
The Green's function $\underline{\mathcal{G}}_{\,c}$ is defined in 
Eq.~\eqref{eq:tmat_green} (i.e., if the potential scattering is present, $V_0\neq0$,
we assume that it is located at the impurity site, ${\bf r}_0$) and $\varphi_{\bf s}=+[-]1$
for ${\bf s}-{\bf r}_0=(\pm 1,0)[(0,\pm 1)]$.
For a more detailed discussion 
see Ref.~\onlinecite{vojtabullaSTM}.

\subsubsection{Bose Kondo Model: $J_{\text{K}}=0$}

For $J_{\text{K}}=0$ the Bose-Fermi Kondo model reduces to the so-called Bose
Kondo model. In the large-$\mathcal{N}$ limit the pseudo-fermion 
self-energy is given by\cite{bkVBS}
\begin{flalign}\label{eq:sigmaf_bos}
& \Sigma_f^{(\text{b})}(i\nu_n)=
    -\frac{\gamma_0^2}{\beta}\sum_{i\omega_n}
    \mathcal{G}_f(i\nu_n+i\omega_n)\,\mathcal{D}_{\phi}^0(i\omega_n,{\bf r}_0),
\end{flalign}
where $\mathcal{D}^0_\phi(\tau,{\bf r}_0)=
-\langle T_\tau^{\phantom{\dagger}} 
\phi_0^{\phantom{\dagger}}(\tau)\phi_0^\dagger(0)\rangle$ is the local Green's function
of the bosonic bath. The self-energy $\Sigma_f^{(\text{b})}(i\nu_n)$ is derived by summing
all diagrams with non-crossing bosonic lines (the so-called \lq\lq rainbow" diagrams) 
which is known as the non-crossing approximation (NCA).\cite{bkVBS,mvIMP}
The self-consistency is obtained using \eqref{eq:fullGf}, \eqref{eq:half_sc},
and \eqref{eq:sigmaf_bos}.

Interestingly, in the Bose Kondo model there is always a non-trivial interaction 
between the impurity and the bosonic bath, i.e., $\Sigma_f^{(\text{b})}(i\nu_n)\neq0$.
At zero temperature and at the bulk quantum critical point ($m=\Delta_{\text{s}}=0$ and 
$\omega_{\bf q}=cq$) the self-consistent solution  yields a power-law behavior 
for the pseudo-fermion spectral function, $\rho_f(\omega)\sim|\omega|^{\epsilon/2-1}$.
This behavior does not correspond to the usual Kondo screening which occurs in the fermionic
Kondo model, but rather reflects non-trivial fluctuations that are usually 
referred to as the bosonic fractional (fluctuating) phase. 
For finite but small $\Delta_{\text{s}}$ there are no bulk spin excitations
at low energies. For $\omega\ll\Delta_{\text{s}}$ the impurity moment is decoupled from
the bath while the case $\omega\gg\Delta_{\text{s}}$ 
is analogous to the case $\Delta_{\text{s}}=0$;
the crossover between these two regimes has been studied in detail in Ref.~\onlinecite{bkVBS}.

\subsubsection{Bose-Fermi Kondo Model}

A generalization of the large-$\mathcal{N}$ approach for the Bose-Fermi Kondo model
cannot be done in a straightforward manner and further approximations are necessary. 
(We note, however, that the large-$\mathcal{N}$ analysis has been 
applied to a {\it multi-channel} version of
the Bose-Fermi Kondo model.\cite{bfkLN})
Guided by the fact that the physics of both the fermionic and the bosonic Kondo model 
is controlled by a saddle point in the $\mathcal{N}\to\infty$
limit, we combine both effects through the 
pseudo-fermion self-energy
\begin{flalign}\label{eq:sigmaf}
& \Sigma_f^{\phantom{\dagger}}(i\nu_n)=
    \Sigma_f^{(\text{f})}(i\nu_n)+\Sigma_f^{(\text{b})}(i\nu_n).
\end{flalign}
The fermionic part obtained within the slave-boson mean-field approach \eqref{eq:sigmaf_fer}
and the bosonic NCA contribution \eqref{eq:sigmaf_bos}
are only formally decoupled. 
Their mutual effect can be easily understood through 
the self-consistency which we present below.
Namely, diagrams containing both the conduction electron and the 
bosonic propagators are included through the full $f$-propagator in 
$\Sigma_f^{(\text{b})}(i\nu_n)$. 
In addition, according to Eq.~\eqref{eq:fullGfc}
the mixed propagator, $\mathcal{G}_{fc}$,
also contains the full pseudo-fermion propagator. Therefore,
the interaction with the bosonic bath entering the mixed propagator
through $\mathcal{G}_{f}$ influences the condensation amplitude, 
i.e., Kondo screening.

The system of equations \eqref{eq:fullGf}-\eqref{eq:sigmaf} can be solved self-consistently
in order to obtain the full $f$-propagator, $v$, and $\lambda_0$ for given
couplings $J_{\text{K}}$ and $\gamma_0$, and for given local Green's functions of the 
fermionic and the bosonic bath, $\mathcal{G}_{c}^0(i\nu_n,{\bf r}_0)$ and 
$\mathcal{D}_\phi^0(i\omega_n,{\bf r}_0)$, respectively; 
the results obtained by numerically solving this self-consistent 
system are presented Sec.~\ref{sec:kondoresp}. 

Before we show our numerical results in the next section, 
we discuss some limiting cases.
For small coupling $J_{\text{K}}$ and
the fermionic bath exponent $r>0$ we expect no Kondo screening, $v=0$,
and that the problem reduces to the Bose Kondo model. On the other hand,
for large $J_{\text{K}}$ and small $\gamma_0$ the pseudo-fermion self-energy
is dominated by its fermionic term and we should recover the results of the fermionic 
pseudogap Kondo model. Consequently, the two interactions of the impurity
with the fermionic and the bosonic bath (i.e., the fermionic and the bosonic
Kondo physics) compete; the same conclusion has been reached within the
renormalization group approach.\cite{bfkSi,BFK1,BFK2} 

For $m=\Delta_{\text{s}}=0$ there is a quantum phase transition
between a fermionic Kondo-dominated phase, for $J_{\text{K}}\gg\gamma_0$, 
and a bosonic fluctuation-dominated phase, for $J_{\text{K}}\ll\gamma_0$.
Note that there is no zero temperature free moment phase as
long as $\gamma_0\neq0$ and $m=\Delta_{\text{s}}=0$;
this is the so-called bosonic fractional moment phase.\cite{bkVBS}
An interesting quantity is $T_{\text{K}}=T_{\text{K}}(J_{\text{K}})$ 
where the Kondo screening sets in. It can be derived using Eq.~\eqref{eq:v_sc}
provided that $v\to0$ and $\lambda_0\to0$. In our calculation we determine the critical
value of the Kondo coupling, $J_{\text{Kc}}$, in the $T_{\text{K}}\to0$ limit.


\section{Results}
\label{sec:results}

In this section we present our numerical results which are obtained 
for a system at zero temperature. 
First we explain the way in which the LDOS of the host SC is modified. Afterwards, 
we show the results for both (i) the pure PS model and 
(ii) the Kondo model with and without the potential scattering at the impurity
site. Finally, we compare them with the existing STM experimental data.

\subsection{Host Local Density of States}
\label{sec:ldos}

As already noted recent STM measurements of the LDOS in cuprates 
have led to numerous studies; they
deal with the possible origin of the nanoscale spatial 
variation\cite{mcelroySCI,nunner,uchida} or 
identify the processes contributing to the tunneling spectra.\cite{pilgram} 
In the present calculation we do not attempt to explain the 
origin of the nanoscale inhomogeneities
but we model the host LDOS 
in a way to get close resemblance to the STM data.\cite{langBSCCO,mcelroy}

The STM experiments have established that the LDOS shows a 
nanoscale spatial modulation with 
two types of regions:\cite{howald,langBSCCO} 
(i) The first type corresponds to domains with small gaps and
well-defined coherence peaks. In this region we assume that the BCS Hamiltonian
\eqref{eq:ham_BCS} describes the bulk superconductor and 
we use the conduction electron Green's function 
defined in Eq.~\eqref{eq:tmat_green} to obtain the LDOS.
(ii) The second type of domains have large gaps 
and very broad gap-edge peaks whose amplitude is relatively low. 
In the following we will refer to these regions as the \lq\lq pseudogaps".

In order to describe the host superconductor in the \lq\lq pseudogap" domains 
we use a rather phenomenological approach. 
We write the conduction electron Green's function as
$[\underline{\tilde{\mathcal{G}}}_{\,c}(i\nu_n,{\bf k})]^{-1}=
 [\underline{\mathcal{G}}_{\,c}^0(i\nu_n,{\bf k})]^{-1}-
 \underline{\Sigma}_{\,c}(i\nu_n,{\bf k})$, where
the bare propagator is the propagator in the pure BCS state.
The main approximation enters through the form of the conduction electron
self-energy, $\underline{\Sigma}_{\,c}(i\nu_n,{\bf k})$, which we do not
calculate but we employ a simple model for it. 
(For a thorough discussion about the conduction electron
self-energy we refer the reader to Ref.~\onlinecite{chubukov1}.)
It is important to note that the self-energy is chosen 
in such a way that the corresponding LDOS obtained from the perturbed
Green's function, $\underline{\tilde{\mathcal{G}}}_{\,c}(i\nu_n,{\bf k})$,
mimics the LDOS measured by the STM on BSCCO.\cite{langBSCCO,mcelroy,mcelroySCI}
As already mentioned, the local Green's function of the host SC is used further 
as an input (i.e., a given) quantity for the $T$-matrix and the 
large-$\mathcal{N}$ calculations.
We assume that the imaginary part of the self-energy is given by
$\Sigma_{11}''(\omega,{\bf k})
\sim-\omega^2/\Delta_0$ for $|\omega|\le\Delta_0$, 
and $\Sigma_{11}''(\omega,{\bf k})\sim-\Delta_0$ otherwise.\cite{note_k}
The $\omega^2$-dependence ensures that the low-energy region of
the LDOS (e.g., for $\omega\ll\Delta_0$)
is not modified in all nanoscale domains
which is in accordance with the STM measurements.\cite{langBSCCO,mcelroy}
The real part of the self-energy is obtained by using the Kramers-Kronig relation.
In addition, for simplicity we assume that 
$\Sigma_{11}=\Sigma_{12}=\Sigma_{21}^\ast=\Sigma_{22}$.

\begin{figure}[h!]
\includegraphics{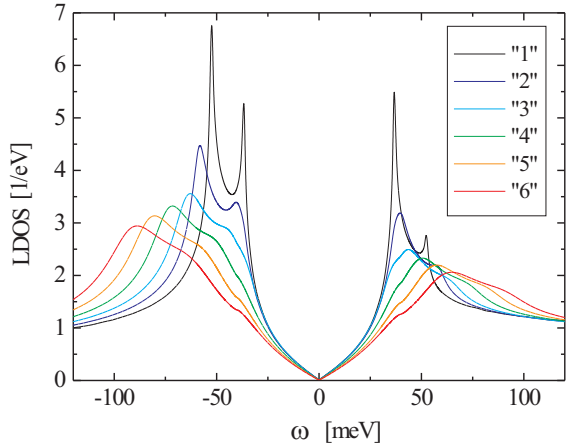}
\vspace*{-6mm}
\caption{(color online) 
The different local densities of states (LDOS) of the bulk superconductor
which are used as inputs for the $T$-matrix and the
large-$\mathcal{N}$ calculations. 
By going from a pure BCS state (denoted by \lq\lq 1") 
towards a \lq\lq pseudogap" state (denoted by \lq\lq 6") 
the gap value increases and the superconducting coherence peaks disappear.
The low-energy region remains unchanged.}
\label{fig:ldos}
\end{figure}

The host LDOS which we use in the present calculation 
are shown in Fig.~\ref{fig:ldos} and are denoted by 
$\lq\lq 1,\ldots,6"$.
The case \lq\lq 1" corresponds to the pure BCS state with well-defined 
coherence peaks and a small gap while in other spectra the gap increases
and the coherence peaks disappear. In this way we 
can model the STM spectra with small (large)
gap regions with (without) SC coherence peaks.
Furthermore, the STM measurements show
that the low-energy region close to the Fermi level 
remains unchanged;\cite{langBSCCO,mcelroy}
this feature is also captured in Fig.~\ref{fig:ldos}.
Note that larger gap values correspond to a lower hole doping level;
we use the gap values $(1.0,\,1.1,\,1.2,\,1.4,\,1.6,\,1.8)\,\Delta_0$
for the spectra $\lq\lq 1,\ldots,6"$, respectively.
By adjusting the chemical potential $\mu$ one obtains
the correct doping; more precisely we fixed
$\mu=-$($130.5$,$\,125$,$\,120$,$\,115$,$\,110$,$\,105$)$\,$meV
for the doping levels
$p=$($19.5$,$\,17.5$,$\,15.9$,$\,14.5$,$\,13.1$,$\,11.8$)$\,\%$.\cite{note_gap}
It is important to emphasize that the doping level does not significantly change
our results and it can be in principle completely ignored.
Nevertheless, we include the different dopings in order to model the experimental
situation as close as possible.

\subsection{Potential Scattering}

We start our discussion with the pure PS model ($J_\text{K}=\gamma_0=0$).
It is known that the delta-function potential scatterer generates
a resonant state in the superconductor LDOS and the resonance can be tuned
towards the Fermi level by increasing $V_0$; the low-energy resonant state is
possible only in unitary limit, i.e., when $V_0$ is the 
largest energy scale in the problem. However, the PS model is not able to 
explain the spatial distribution of the resonance  which is seen in 
STM experiments.\cite{panBSCCO,notefilter}
These results are summarized, for instance, in 
Fig.~1 of Ref.~\onlinecite{andreev}.

\begin{figure}[h!]
\includegraphics{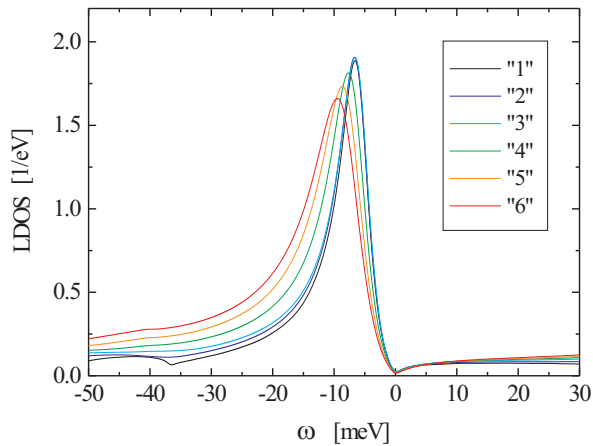}
\vspace*{-6mm}
\caption{(color online) LDOS at the impurity site in the PS model for
$V_0=0.2$\,eV and different host LDOS. A resonance appears at a
low-energy scale $\sim\mathcal{O}(\text{meV})$; it is slightly suppressed and moved away 
from the Fermi level by modifying the superconducting host,
 i.e., by going from \lq\lq 1" to \lq\lq 6" in Fig.~\ref{fig:ldos}.}
\label{fig:PS}
\end{figure}

Here, we consider a non-magnetic impurity as a pure potential scatterer placed 
in different regions of the $d$-wave superconductor; 
those regions are characterized by the 
corresponding host LDOS (shown in Fig.~\ref{fig:ldos}). 
In Fig.~\ref{fig:PS} we show the LDOS in a $d$-wave SC at the
impurity site obtained within the $T$-matrix approach for a given $V_0$
and different host LDOS. The potential scattering strength is 
chosen such that one gets a resonance in the low-energy region.

Modifying the pure BCS state towards the 
\lq\lq pseudogap" state does not change the resonance significantly.
The resonance is slightly moved away from the Fermi level
and its amplitude is somewhat suppressed. We conclude that if the 
Zn resonance in BSCCO were of the PS type then it 
would be present in all regions (both small- and large-gap regions).
This result is not surprising if we recall that in the $T$-matrix
approach only the low-energy part of the host LDOS
can influence the resonance close to the Fermi level.
According to the STM experiments the low-energy part of the host LDOS 
remains unchanged (as in Fig.~\ref{fig:ldos}) so that the Zn resonance
in the PS model survives in all regions 
which is in disagreement with the STM data.
Note that for larger $V_0$ (i.e., for lower energy of the resonant state)
the change in the position and the amplitude of the resonance state 
is even less pronounced.

\subsection{Kondo Response}
\label{sec:kondoresp}

Here we present numerical results for the Bose-Fermi Kondo
model obtained within the large-$\mathcal{N}$ approach.

\subsubsection{Impurity Spectral Function}

\begin{figure}[h!]
\includegraphics{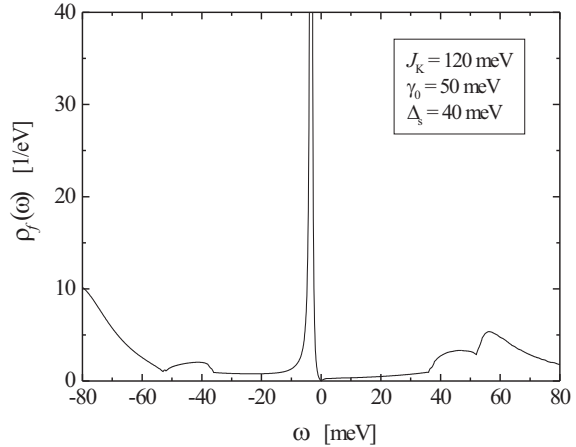}
\vspace*{-6mm}
\caption{A typical impurity spectrum in the Kondo
screened phase for given fermionic and bosonic
coupling constants and for a given spin gap. The scattering
potential is $V_0=0.2$, and the SC
host is in the pure BCS state (\lq\lq 1"). 
The impurity spectral function
develops a peaked structure at a 
characteristic Kondo energy scale $T^\ast$.}
\label{fig:rho_f}
\end{figure}

In Fig.~\ref{fig:rho_f} the typical impurity spectrum 
in the Kondo screened phase ($v\neq0$) is shown; it is obtained
by solving the self-consistent system of equations \eqref{eq:fullGf}-\eqref{eq:sigmaf}
presented above.
In the Kondo screened phase the impurity spectral function, $\rho_f(\omega)$, develops a 
pronounced peak at small but finite energy, 
typically of the order of $T^\ast\sim\mathcal{O}(\text{meV})$. 
Such a peaked feature does not develop in the local moment phase where the
Kondo screening is completely suppressed and $v=0$.

In the present large-$\mathcal{N}$ analysis $T_{\text{K}}$ and $T^\ast$ are not equivalent,
i.e., they do not behave in the same manner upon approaching the transition. 
Nevertheless, $T^\ast$ can be used to characterize Kondo 
screening since (i) the local magnetic susceptibility 
changes its behavior at $T^\ast$ and (ii) the peaked structure at $T^\ast$ also appears in the 
conduction electron $T$-matrix as seen both in the large-$\mathcal{N}$ 
and the NRG calculations.\cite{cassanellofradkin, polkovnikovSTM, vojtabullaSTM} 
(Note, however, that in the NRG calculations the characteristic energy scale, 
$T^\ast$, can be identified with the Kondo temperature up to a numerical 
factor of order unity.\cite{polkovnikovSTM, vojtabullaSTM})
In the present work we perform a zero temperature calculation so 
that $T^\ast$ serves as a natural choice to characterize the Kondo screening. 
Furthermore, we associate the peaked structure in the impurity spectral function 
with the Zn resonance in the STM data.\cite{panBSCCO}
In other words, the Zn resonance is likely caused by a Kondo-like 
behavior of the induced effective magnetic moment 
with the Kondo temperature as a natural low-energy scale.

As already noted the resonance peak seen by STM does not decay monotonically with
distance from the Zn site but rather oscillates, producing local minima and maxima
in the DOS map.  
Remarkably, the four nearest neighbor Cu sites around the impurity have no DOS 
local maxima associated with them (in contrast to the next-nearest
and next-next-nearest neighbor Cu sites).\cite{panBSCCO} 
This non-monotonic spatial dependence cannot be
naturally explained within the PS model. 
On the other hand, the four-site Kondo model for an impurity in a
$d$-wave superconductor can capture this feature.\cite{polkovnikovSTM,vojtabullaSTM}

\subsubsection{Characteristic Kondo Energy Scale}

\begin{figure}[h!]
\includegraphics{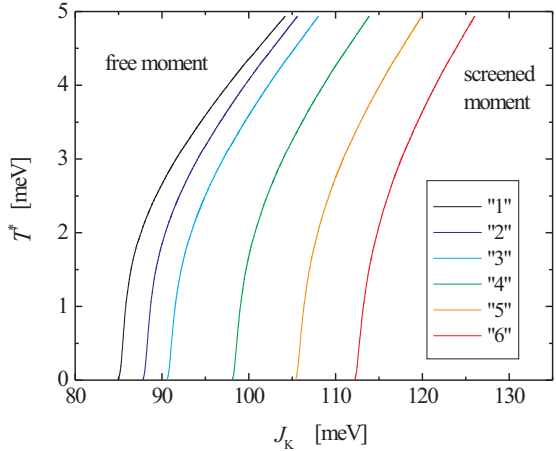}
\vspace*{-6mm}
\caption{(color online) 
The characteristic energy scale as a function
of the Kondo coupling for different host regions (the 
corresponding LDOS are given in Fig.~\ref{fig:ldos}).
The potential scattering and the bosonic bath are absent:
$V_0=0$, $\gamma_0=0$.
The pure BCS state (small-gap domain) has 
lowest $J_{\text{Kc}}$.}
\label{fig:TK_JK}
\end{figure}

Here we analyze how the characteristic energy scale changes with
the Kondo coupling when the impurity is placed in 
different regions of the SC host. 
In Fig.~\ref{fig:TK_JK} we show how $T^\ast$ depends on the Kondo coupling
when a host LDOS is modified from the pure BCS to the
\lq\lq pseudogap" case (the corresponding LDOS 
are shown in Fig.~\ref{fig:ldos}).
We see that a larger value of the Kondo coupling is required
in order to screen the impurity moment in the large-gap regions
(i.e., the large-gap regions have larger critical Kondo coupling, 
$J_\text{Kc}$, below which the moment is free and above which 
the moment is screened).
In particular, for certain values of $J_\text{K}$ the screening is possible only
in the small-gap regions and not possible in the large-gap regions.
In other words, the Kondo-like impurity resonance
is suppressed by going from the pure BCS to the \lq\lq pseudogap" state;
this is an important result which is supported by the experimental data.\cite{langBSCCO} 
This fact cannot be explained within the PS model but naturally 
arises in the Kondo model.

Note that the values of $T^\ast\sim\mathcal{O}(\text{meV})$ 
also agree well with the Kondo temperatures 
estimated in NMR experiments $\sim\mathcal{O}(10\,\text{K})$.\cite{bobroff}
In Ref.~\onlinecite{vojtabullaSTM} it has already been
shown that the presence of weak to moderate potential scattering 
does not change the results of the four-site fermionic Kondo model. 
In contrast, large values of the scattering potential induce a large DOS on 
the neighboring sites of the Zn impurity.
As a consequence the critical Kondo coupling is strongly reduced and the
Kondo scale is increased up to temperatures $\sim\mathcal{O}(100\,\text{K})$, 
which is in disagreement with the NMR data.\cite{bobroff} 
On the other hand, the Kondo temperature is suppressed in the presence 
of the collective bosonic modes which are included 
in the Bose-Fermi Kondo model.
In what follows we present the phase diagram of the Bose-Fermi
Kondo model obtained using the large-$\mathcal{N}$ approach
and also show the suppression of the Kondo screening in the 
presence of the collective bosonic modes.

\subsubsection{The Phase Diagram}

\begin{figure}[h!]
\includegraphics{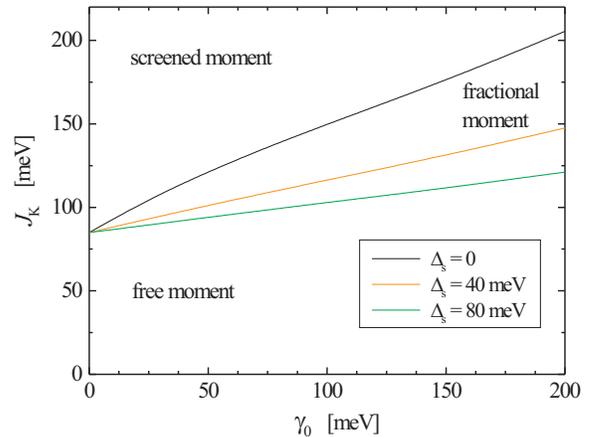}
\vspace*{-6mm}
\caption{(color online) 
The projection of the phase diagram onto the $J_{\text{K}}$-$\gamma_0$ plane;
the critical coupling $J_{\text{Kc}}$ is determined
for different spin-gap values and the host SC in the pure BCS state (\lq\lq 1"). 
For $\Delta_{\text{s}}=0$ and 
$\gamma_0=0$ ($\gamma_0\neq0$) there is a transition between the free 
(fractional) moment phase and the Kondo screened phase.
For $\Delta_{\text{s}}\neq0$ there is only a transition between 
the screened and the free moment phase. The potential scattering
is absent, $V_0=0$.
}
\label{fig:JkGamma0_mass}
\end{figure}

The projection of the phase diagram onto the $J_\text{K}$-$\gamma_0$ plane
is shown in Fig.~\ref{fig:JkGamma0_mass} where the critical coupling $J_\text{Kc}$
is determined for different values of the spin gap and the host SC in the pure BCS state (\lq\lq 1").
For $\gamma_0=0$ we recover a fermionic pseudogap Kondo model with
a finite value of $J_\text{Kc}$ which is, as expected, independent of the spin
gap; below $J_\text{Kc}$ the impurity moment is free. 
For $\gamma_0\neq0$ and $J_\text{K}<J_\text{Kc}$ we distinguish 
two different cases: 
(i) For the gapless bosonic bath, $\Delta_{\text{s}}=0$, the impurity 
moment is not completely free but rather placed into a bosonic fractional 
(fluctuating) phase.\cite{bkVBS} 
(ii) A coupling to a gapped bosonic bath, $\Delta_{\text{s}}\neq0$,
leaves the impurity completely decoupled from both baths.
Furthermore, fermionic and bosonic Kondo physics compete since
the larger $\gamma_0$ values require larger $J_\text{Kc}$ in order to
screen the moment. (Note that the same conclusion was reached within the 
RG approach for the Bose-Fermi Kondo model with 
$\Delta_{\text{s}}=0$.\cite{BFK1,BFK2})
Also, for larger spin-gap values the bosonic bath is
less effective in destroying the fermionic Kondo screening. (This fact 
was proposed earlier to explain the NMR data showing a 
suppression of the Kondo screening by underdoping.\cite{BFK1}) 

\begin{figure}[h!]
\includegraphics{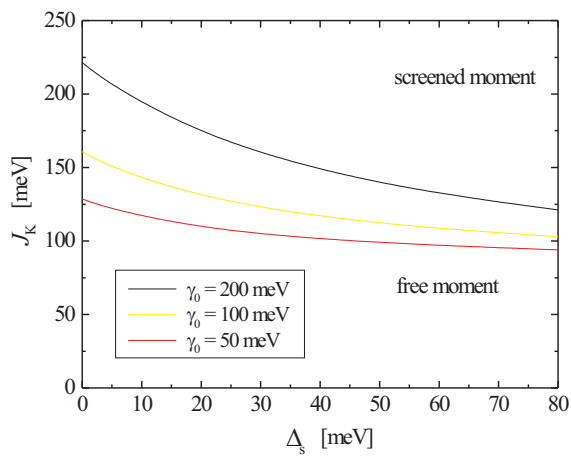}
\vspace*{-6mm}
\caption{(color online) 
The projection of the phase diagram onto the 
$J_{\text{K}}$-$\Delta_{\text{s}}$ plane;
the critical coupling $J_{\text{Kc}}$ is determined
for different values of the bosonic coupling and the host SC in
the pure BCS state (\lq\lq 1").
For $\Delta_{\text{s}}=0$ ($\Delta_{\text{s}}\neq0$)
there is a transition between the fractional (free) 
moment phase and the Kondo screened phase. 
The potential scattering is absent: $V_0=0$.
}
\label{fig:Jkmass_Gamma0}
\end{figure}

In Fig.~\ref{fig:Jkmass_Gamma0} we show the projection of the phase 
diagram onto the $J_{\text{K}}$-$\Delta_{\text{s}}$ plane. 
The critical Kondo coupling is determined for different 
values of the bosonic coupling. For $\Delta_{\text{s}}=0$ ($\Delta_{\text{s}}\neq0$)
there is a transition between the fractional (free) moment phase 
for $J_{\text{K}}<J_{\text{Kc}}$, and the Kondo screened phase
for $J_{\text{K}}>J_{\text{Kc}}$. By increasing the spin gap 
the number of low-energy bosonic excitations is reduced which makes the bosonic bath
less effective in destroying the Kondo screening
and as a consequence $J_{\text{Kc}}$ decreases.
The bosonic spin gap is doping dependent and 
decreases with underdoping.\cite{fong}
As a consequence, one expects that the bosonic bath 
stronger suppresses the Kondo screening in the region with 
lower hole doping which corresponds to the region with
the larger-gap in LDOS.

As mentioned above in the presence of the potential scattering the 
DOS on the neighboring sites of the Zn impurity increases 
which results in a decrease of the critical Kondo coupling.
For example, for $\gamma_0=50$\,meV and $\Delta_{\text{s}}=40$\,meV and the host SC in
the region \lq\lq 1", we have 
$J_{\text{Kc}}=(101.3,\,100.4,\,97.3)$\,meV for $V_0=(0,\,0.2,\,0.5)$\,eV, 
respectively.

\subsubsection{Suppression of the Kondo Resonance}

\begin{figure}[h!]
\includegraphics{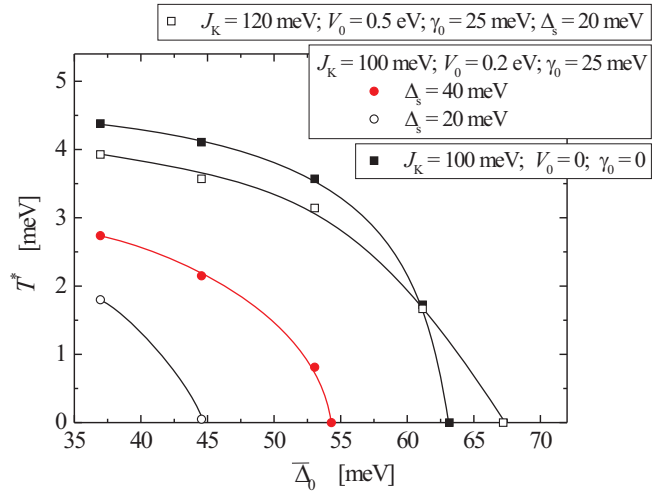}
\vspace*{-6mm}
\caption{(color online)
The characteristic energy scale as a function of the 
average energy gap:
$J_{\text{K}}=120$\,meV, $V_0=0.5$\,eV, $\gamma_0=25$\,meV,
$\Delta_{\text{s}}=20$\,meV (open squares);
$J_{\text{K}}=100$\,meV, $V_0=0.2$\,eV, $\gamma_0=25$\,meV, 
$\Delta_{\text{s}}=40$\,meV (solid circles) and
$\Delta_{\text{s}}=20$\,meV (open circles);
$J_{\text{K}}=100$\,meV, $V_0=0$, $\gamma_0=0$ (solid squares);
the lines are guides to the eye.
By going from the small-gap region to the
large-gap region the Kondo screening of the impurity moment
vanishes.
}
\label{fig:TK_Delta0}
\end{figure}

Here we discuss how the characteristic energy scale changes by
modifying the superconducting host. Figure \ref{fig:TK_Delta0}
shows the Kondo scale, $T^\ast$, as a function of an average energy gap 
$\bar{\Delta}_0$. The average energy gap is defined as 
$\bar{\Delta}_{0}=(\Delta_-+\Delta_+)/2$
where the quantity $\Delta_-$ ($\Delta_+$) is the energy of the first 
peak below (above) the Fermi level in the host LDOS without impurity.
For the pure BCS state $\bar{\Delta}_0$ 
corresponds to a superconducting gap $\Delta_0$ while in the \lq\lq pseudogap"
state one can estimate it from the LDOS shown in Fig.~\ref{fig:ldos}.

All sets of data points in Fig.~\ref{fig:TK_Delta0} indicate that the Kondo 
effect is more suppressed in the regions with a larger energy gap.
Furthermore, the suppression of the Kondo screening is a monothonic 
function of the average energy gap and it is, in principle, possible to identify the
maximum value of the energy gap above which the impurity resonance cannot
appear in the LDOS;
this agrees with the STM measurements.\cite{langBSCCO}
(Note that the amplitude of the Andreev resonant state in 
Ref.~\onlinecite{andreev} is a non-monotonic function of the energy gap;
the largest resonance amplitude one obtains for intermediate gap values.)
It is important to note that the Kondo energy scale $\sim\mathcal{O}(\text{meV})$
can be tuned to zero by simply modifying the host LDOS; 
we have shown that this is not possible in the PS model (Fig.~\ref{fig:PS}).
On the other hand, in the Kondo model the low-energy region of
the bath LDOS is important for the Kondo effect, but through the self-consistent procedure
also the LDOS away from the Fermi level influences the Kondo screening. 
Since the high-energy part of the bath LDOS differs significantly in 
small- and large-gap regions, it is natural to expect that the Kondo resonance
will be affected when the host LDOS is modified.

Figure \ref{fig:TK_Delta0} also suggests that the presence of the bosonic bath
additionally suppresses the Kondo screening. 
For the spin gap, $\Delta_{\text{s}}$, we use 
a constant value regardless of the specific host LDOS,
i.e., regardless of doping level. 
We could, however, associate
different spin gaps to different host LDOS since the spin gap is 
doping dependent;\cite{fong} the underdoped regions 
(i.e., the regions with a larger energy gap $\bar{\Delta}_0$) 
have smaller $\Delta_{\text{s}}$.
In this case, the suppression of $T^\ast$ with 
$\bar{\Delta}_0$ would be even more pronounced.


\section{Conclusions}
\label{sec:concl}

Motivated by the question about the origin of the impurity resonance in 
$d$-wave SC we have compared two frequently discussed scenarios, 
i.e., the potential scattering model and the Kondo model.
We have investigated how the STM impurity resonant state
is influenced when the SC host is chosen in such a way that
it resembles the nanoscale electronic inhomogeneities seen by STM.\cite{langBSCCO}
(i) If the impurity atom is assumed to be a pure potential scatterer
we have shown within the $T$-matrix approach that the impurity resonance 
appears as a robust feature. The resonant state
is rather insensitive to changes of the SC host. It is 
present in both small- and large-gap regions which is 
in disagreement with the experimental data.
(ii) The suppression of the impurity resonance in the large-gap
domains observed by STM can be naturally explained by assuming that the impurity atom 
gives rise to Kondo-type physics.
We have used the Bose-Fermi Kondo model and the large-$\mathcal{N}$ method
to show that the characteristic Kondo energy scale is affected by changing 
the host properties (i.e., by changing the corresponding host LDOS).
The impurity resonance resulting from the Kondo screening 
can appear in the host small-gap regions and can be completely 
suppressed in the large-gap regions; 
this is in agreement with the STM experimental
data for Zn-doped \bscco. 
Therefore, we conclude that the Kondo spin dynamics
of the impurity moment is the origin of the impurity resonant state
and that the pure potential scattering model is not sufficient
to capture the physics of the non-magnetic impurity 
in $d$-wave superconductors.


\acknowledgments

We thank A. V. Chubukov and
M. Vojta for fruitful discussions,
W. Metzner for a careful reading of the manuscript, 
and especially H. Yamase for a critical 
reading and numerous useful comments.



\begin{thebibliography}{}
\bibitem{notefilter} To obtain the results which are in qualitative 
              agreement with the STM data one must include the so-called
              filter effect.\cite{martinfilter,balatskyIMP}
\bibitem{notecoupling} It has been shown that in the low-energy limit
              a possible coupling between the fermionic and the bosonic bath is forbidden 
              by momentum conservation.\cite{bkVBS}
\bibitem{note_k} In principle, we could introduce a ${\bf k}$-dependent factor
                 in the expression for the self-energy
                 to account for the difference in the nodal and antinodal
                 regions; a suitable choice would be
                 $|\cos{k_x}-\cos{k_y}|$ or $(\cos{k_x}-\cos{k_y})^2$. 
                 However, after the momentum summation this change 
                 would not affect the low-energy local DOS  
                 because the $\omega^2$-dependence in the self-energy 
                 is subdominant in comparison with the linearly
                 vanishing DOS for the bare $d$-wave superconductor.
\bibitem{note_gap} Experimentally it is not possible to 
                 determine the local hole doping and associate it with the 
		 certain gap value.\cite{mcelroy} Nevertheless, one can
		 measure the bulk hole doping and then associate it 
		 with the spatially averaged gap value, $\bar{\Delta}$.
\end{thebibliography}
\end{document}